\documentclass[preprint2]{aastex}

\shorttitle{NIR AO Spectroscopy of Binary Brown Dwarf} 
\shortauthors{Goto et al.}

\begin{document}


\title{Near-Infrared Adaptive Optics Spectroscopy of \\ Binary Brown
Dwarf HD~130948B and C\footnote{Based on data collected at Subaru
Telescope, which is operated by the National Astronomical Observatory
of Japan.}}

\author{Miwa Goto\footnote{Visiting astronomer at the Institute for
Astronomy, University of Hawaii}, Naoto Kobayashi, Hiroshi Terada, \\
Wolfgang Gaessler, Tomio Kanzawa} \affil{Subaru Telescope, 650 North
A`ohoku Place, Hilo, HI 96720} \email{mgoto@duke.ifa.hawaii.edu}

\author{Hideki Takami, Naruhisa Takato, Yutaka Hayano, 
\\ Yukiko Kamata, Masanori Iye} 
\affil{National Astronomical Observatory of
Japan, Mitaka, Tokyo 181-8588, Japan}

\author{D. J. Saint-Jacques} 
\affil{D\'epartment de physique, Universit\'e de Montr\'eal,
Montr\'eal (Qu\'ebec), Canada, H3C 3J7} 

\and

\author{A. T. Tokunaga, D. Potter, M. Cushing}
\affil{Institute for Astronomy, University of Hawaii, 
    Woodlawn Dr., Honolulu, HI 96822}

\begin{abstract}
We present near-infrared spectroscopy of low-mass companions in a
nearby triple system HD~130948 (Gliese~564, HR~5534). Adaptive optics
on the Subaru Telescope allowed spectroscopy of the individual
components of the 0\farcs13 binary system. Based on a direct
comparison with a series of template spectra, we determined the
spectral types of HD~130948B and C to be L4 $\pm$ 1. If we take the
young age of the primary star into account (0.3--0.8~Gyr), HD~130948B
and C most likely are a binary brown dwarf system.

\end{abstract}


\section{Introduction} 
HD~130948 is a triple system discovered by \citet{pot01} using the
University of Hawaii adaptive optics system (H\=ok\=upa`a) in the
course of an imaging survey of low-mass companions to nearby young
stars using the Gemini North telescope. The HD~130948 (Gliese~564,
HR~5534) system consists of a G2V primary star and a binary companion
at a separation of 2\farcs63 from the primary. The binary companions
are separated by 0\farcs134, or 2.4~AU at a distance of 17.9~pc
determined by {\it Hipparcos}. \citet{pot01} confirmed the common
motion of the members of the triple system with a baseline of 7 months
to reject the chance projection of background sources. By using the AO
system on the Subaru Telescope, we obtained individual spectra of each
component to confirm the substellar nature of the faint companions,
HD~130948B and C.

\section{Observations and Data Reduction}
The spectroscopic observation was made on UT 2001 May 3 using the
Infrared Camera and Spectrograph (IRCS; Tokunaga et al. 1998;
Kobayashi et al. 2000) with the 8.2~m Subaru Telescope in conjunction
with its adaptive optics (AO) system \citep{tak98,gae01}. The Subaru
AO system, which is equipped with a 36-element curvature sensor, is
installed at the front end of the telescope Cassegrain port. A
medium-resolution grism was used with a 0\farcs10 slit in the 22 mas
camera section of the IRCS to provide spectra with a resolving power
of 800--1000 in the $H$ and $K$ bands. HD 130948A was used as the
reference source for the AO system. The slit was put along the
position angle of HD~130948B and C to cover them simultaneously. The
spectra were recorded by nodding the telescope by 1\arcsec~along the
slit to subtract the sky emission. The total on-source integration
time was 1200 s in both the $H$ and $K$ bands. A nearby B8V star
HD~164352 was observed as a spectroscopic standard at similar
airmass. The standard star is bright enough to give nearly the same AO
correction as for HD 130948A. The spectroscopic flat field was
obtained at the end of the night with a halogen lamp. The seeing was
0\farcs4--0\farcs5 at 2.2~\micron~throughout the observing period.

Because of the close proximity and the large difference in luminosity
between the primary and companions, the sky background in the spectrum
is affected by the gradient in the wing of the point-spread function
(PSF) of the bright primary. After a pair subtraction of the sky the
residual background slope was removed by interpolating the background
on both sides of the companion.

We obtained the one-dimensional spectra using the IRAF\footnote{IRAF
is distributed by the National Optical Astronomy Observatories, which
are operated by the Association of Universities for Research in
Astronomy, Inc., under cooperative agreement with the National Science
Foundation.} aperture extraction package after flat-fielding and
bad-pixel correction. Special attention was paid to the adjustment of
the aperture width.  Figure \ref{prof1} shows crosscuts of the
observed spectrogram in the spatial direction with an overlay of the
standard star profile normalized at the peak of each object.  The FWHM
of a standard star is as sharp as 0\farcs1 and 0\farcs08 in the $H$
and $K$ bands, respectively. However, the intensity of the first
diffraction ring is significant. The separation between the centroid
and the peak of the first diffraction ring for an ideal
diffraction-limited image of the 8.2~m circular opening should be
0\farcs07 and 0\farcs09 in the $H$ and $K$ bands, respectively. This
is similar to the angular separation of 0\farcs13 between HD~130948B
and C. Consequently, the energy of HD~130948B in the first ring as
well as the halo overlaps with HD~130948C. This may produce similar
spectral signatures in both components that may be misleading.

We calculated the contamination expected with the standard star
PSF. The flux contamination of C by B within the 3 pixel (0\farcs067)
aperture width is 25\% and 22\% in the $H$ and $K$ bands,
respectively. The contamination of B by C is $<$20\%. The extraction
aperture width of the standard star is set to the same as that of the
objects. To provide the correct weight to the optimal extraction
method \citep{hor86}, the statistical error in the 2D spectrogram
image is calculated by combining the readout noise and the shot noise
from the sky emission and PSF spillover from the primary. The
wavelength calibration was performed by fitting about 20 emission
lines of the argon arc lamp of the Subaru Telescope calibration unit
with a linear function.

The intrinsic stellar lines in the standard star spectra, which are
mostly atomic hydrogen lines, were difficult to remove because of
their superposition with the telluric absorption lines. Only six of
them at 1.588~\micron~(Br10), 1.641~\micron~(Br8),
1.681~\micron~(Br7), 1.737~\micron~(Br6), 2.166~\micron~(Br$\gamma$),
and 2.374~\micron~(Pf20) were fit with a Lorentzian profile and
subtracted before division.
The small discrepancy in the airmass between the standard and the
object was corrected by rescaling the standard star spectra according
to Beer's law. Rough flux calibration relative to the standard star
was made in each band by assuming the intrinsic spectra of a B8V star
is represented by a Planck function of $T_{\rm eff} = $ 10700~K
\citep{tok00}.

\section{Result and Discussion}
\subsection{Spectral Type of HD~130948B and C}

To determine the spectral types of HD~130948B and C, we compared the
object spectra with template spectra of low-mass stars published by
\citet{rei01b}, \citet{geb01}, \citet{leg00}, and \citet{leg01a}. We
found none of the template spectra match well with the object
spectra. The continua of the object spectra were too steep to be a
low-mass star, though the low-mass nature is apparent in the strong
absorption of H$_2$O at both ends of the $H$ and $K$ bands. We
concluded that the continuum slope was affected by the AO
correction. Not only a little offset from a narrow slit, but also
slight difference in seeing can result in the variation of the
continuum slope because of the sensitive wavelength dependency of the
encircled energy on the AO performance (M. Goto et al., in
preparation). The variation of the continuum slope from exposure to
exposure was sometimes as much as 10\% per 0.1~$\mu$m. Most of the
time the difference was well represented by a simple linear function.

Because of the uncertainty of the continuum slope, we had to correct
the object spectra by multiplying a linear function before comparison
with templates. The strength of H$_2$O absorption increases with
spectral type, and the spectral change between any two spectral types
is not a linear function. In other words, we cannot convert the
spectrum of an L1 star to an L8 star by just multiplying a linear
function. Therefore, in principle, we can determine the spectral type
of an object uniquely by matching both the H$_2$O absorption with a
template spectrum, even if we have a linear continuum slope
uncertainty.

First, we assumed HD~130948B (or C) is a certain spectral type, and
estimated how much linear continuum correction is required to best
match the object spectrum with the template spectrum of the assumed
spectral type. We calculated the cross-correlation between the object
spectrum and the template spectrum as a function of the steepness of
the applied linear function correction. The slope of the function is
determined so that the cross-correlation between the corrected object
spectrum and the template spectrum is highest. We repeated this
procedure for each spectral type from L1 to L8.

Second, we investigated how well it matched each spectral type from L1
to L8. The value of the cross correlation calculated above peaks at
L3--L5 in the $H$ band and at L1--L5 in the $K$ band. 
Considering the simultaneous match in the $H$ and $K$ bands, we
conclude both HD~130498B and C agree best with the template for L4
$\pm$ 1. This is consistent with the spectral type of L2 $\pm$ 2
determined independently by \citet{pot01}.


The final object spectra after correction are shown in Figure
\ref{spec1} with overlay of the best-matched template spectra, 2MASSW
J0036+18 (L4). The spectra were binned by 2 pixels along the
dispersion direction.  The spectra of HD~130948B and C turn out to be
very similar to each other in detail and are almost twins, except that
HD~130948B is 0.2--0.3 ~mag brighter than C. The deep absorption
bandhead of the CO molecule at 2.3--2.4~$\mu$m is conspicuous. The
presence of absorption lines of \ion{Na}{1} (2.209~$\mu$m) and
\ion{K}{1} (1.516~$\mu$m) supports the idea that these are low mass
stars. The absorption band of CH$_4$ is not seen.  This is in good
agreement with the conclusion that HD~130948B and C are not later than
mid L-type stars. A series of the ``unidentified'' absorption features
reported by \citet{rei01b} at 1.58, 1.613, and 1.627~$\mu$m can be all
attributed to the FeH molecule \citep{wal01}.
\ion{Ca}{1} lines noted by \citet{jon94} and \citet{tin93} are marked
in the figures.

\subsection{Mass of HD~130948B and C}

The range of spectral types derived above corresponds to an effective
temperature of $T_{\rm eff} = 1900 \pm 75$ \citep{leg01a}. On the
basis of moderate X-ray activity, a fast rotation period ($\sim$ 7.8
days), and photospheric lithium abundance, the age of the primary
should be 0.6 $\pm$ 0.2~Gyr \citep{gai98,gai00}. On the other hand
K. Fuhrmann (2001, private communication) finds that HD~130948A may be
a member of the Ursa Major Stream, which has estimated age of 0.3 to
0.65~Gyr \citep{gia79,sod93,pal86,che97}.

Figure \ref{bara1} is a reproduction of the evolutionary tracks of
various low-mass stars of solar metallicity calculated by
\citet{bar98}. Regardless of the uncertainty in the age determination
of the HD~130948 system, the estimated masses for HD~130948B and C are
under the sustainable hydrogen burning limit, 0.075~M$_\sun$ for solar
metallicity, for both companions, and the most likely mass of the
object is 0.040--0.065~M$_\sun$.


\section{Summary}
We have presented moderate-resolution near-infrared spectra of the
close binary companions recently discovered near HD~130948. The
general spectral features of the companions, HD~130948B and C, are
very similar. Based on a direct comparison with template spectra of
low-mass stars, we have determined the spectral type of HD~130948B and
C is L4 $\pm$ 1. This is consistent with the result of \citet{pot01},
who found a spectral type of L2 $\pm$ 2 for the binary companions. If
we take the young age of the primary star (0.3--0.8~Gyr) into account,
then HD~130948B and C are binary brown dwarfs.

\acknowledgments We thank T. Geballe, S. Leggett, I. Neill Reid, and
their collaborators for making the template spectra available in
machine readable form for use in Figure \ref{spec1}. We also thank
I. Baraffe and collaborators for making the tabular data available at
their ftp site for use in Figure \ref{bara1}. We thank all the staff
and crew of the Subaru Telescope and NAOJ for their valuable
assistance obtaining these data and their continuous support for IRCS
construction. M. Goto is supported by a Japan Society for the
Promotion of Science fellowship.

\clearpage


 \figurenum{1} \figcaption{Cross-section of the spectrogram across the
 slit length in the $H$ and $K$ bands with an overlay showing the
 spatial profile of the standard star. The standard star profile is
 registered to the position of each component and normalized at the
 peak of the object profile. The shaded areas represent the size of
 the aperture that the one-dimensional spectra are extracted
 from. \label{prof1}}






\figurenum{2} \figcaption{$H$- and $K$-band spectra of HD~130948B and
C normalized to the flux of HD~130948B at 1.65 and 2.20~$\mu$m,
respectively. The overlay is the best-matched template spectrum,
2MASSW J0036+18 (L4). \label{spec1}}

\figurenum{3} 
\figcaption{The tracks of effective temperature for low-mass stars as
a function of age \citep{bar98}. The effective temperature range for
HD~130948B and C is indicated by the rectangular enclosure, along with
the age determination uncertainty. \label{bara1}}

\clearpage


 \figurenum{1}
 \begin{figure}
 \plottwo{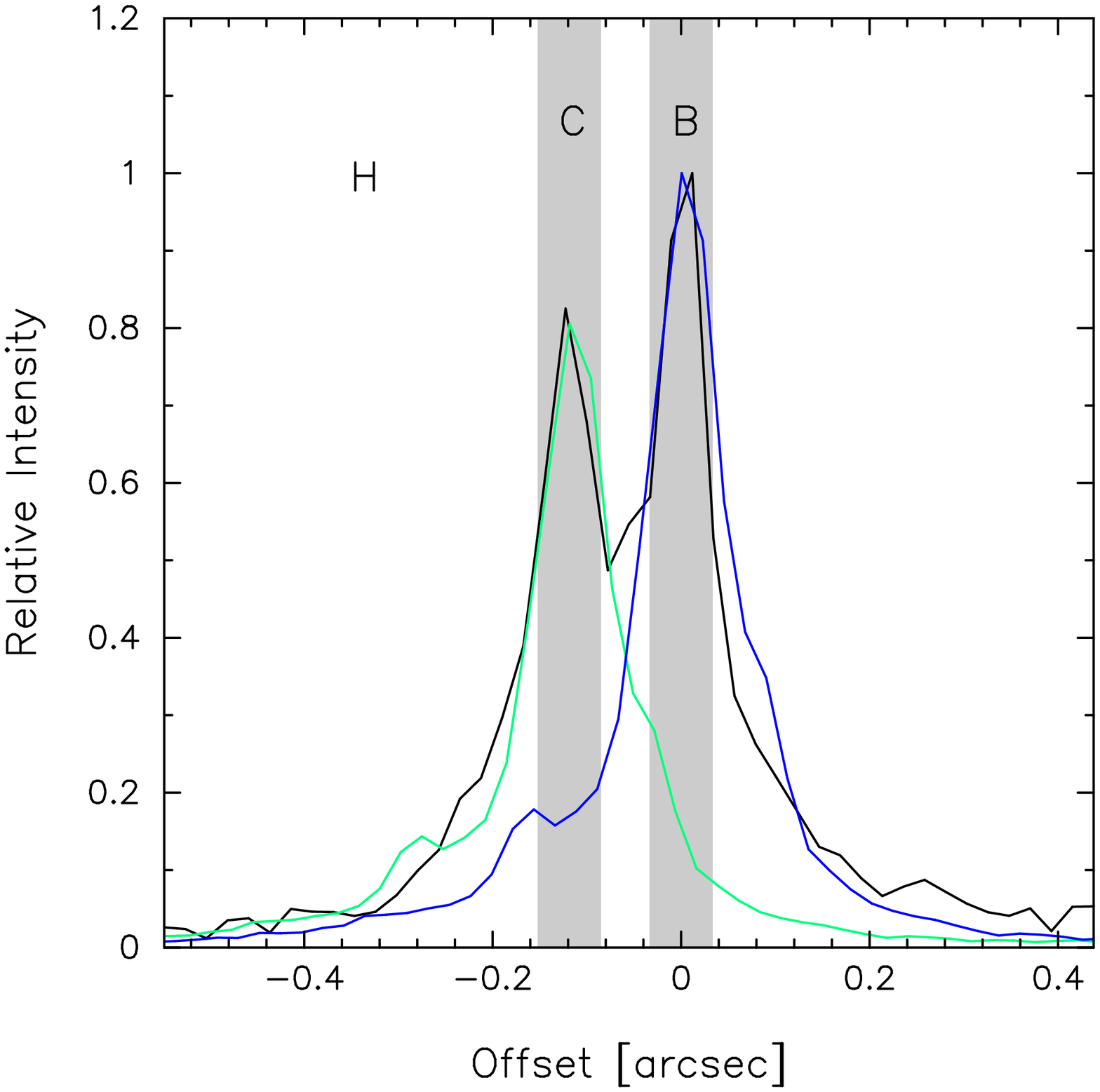}{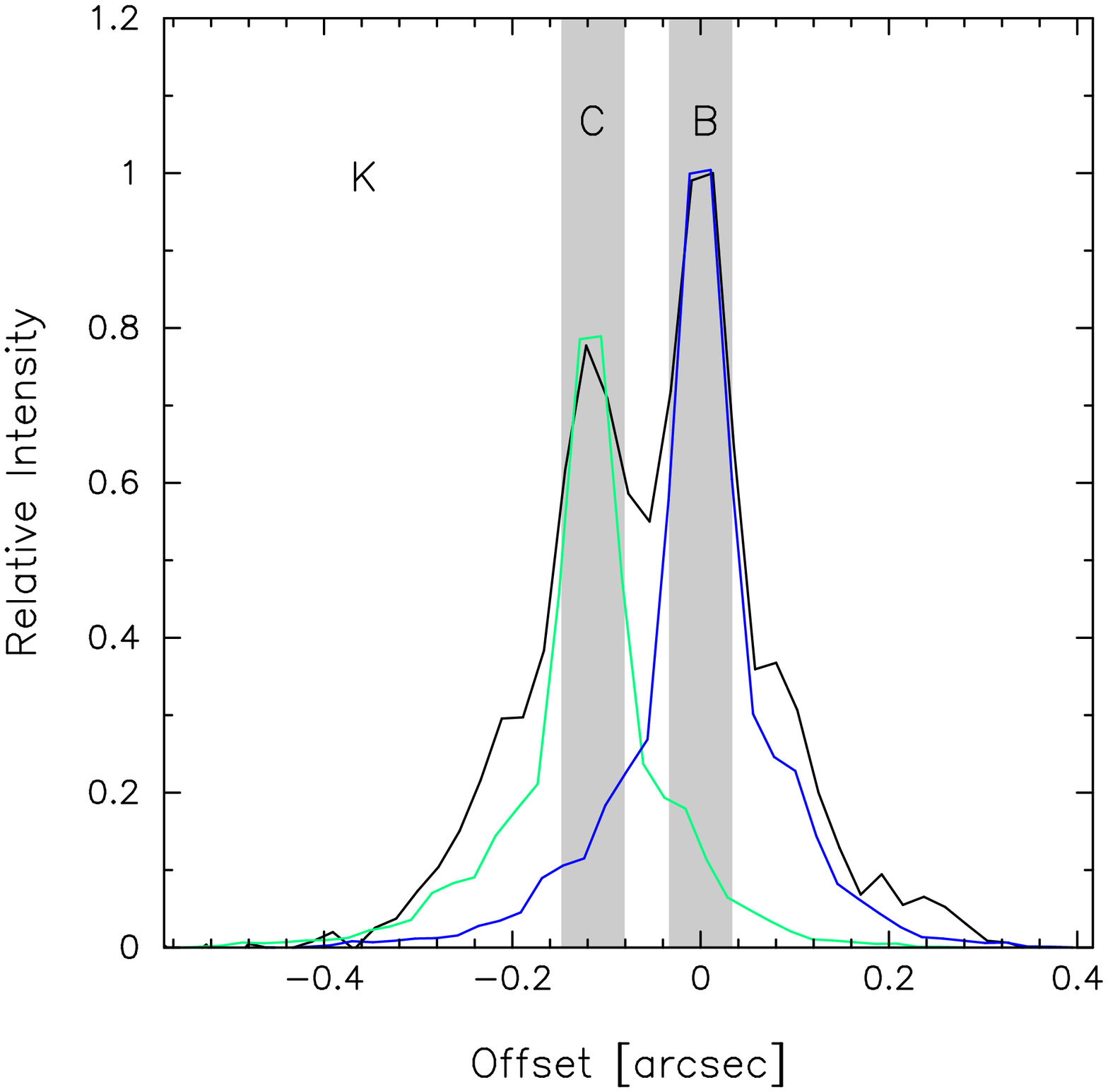}
 \caption{}
 \end{figure}





\figurenum{2}
\begin{figure}
\plotone{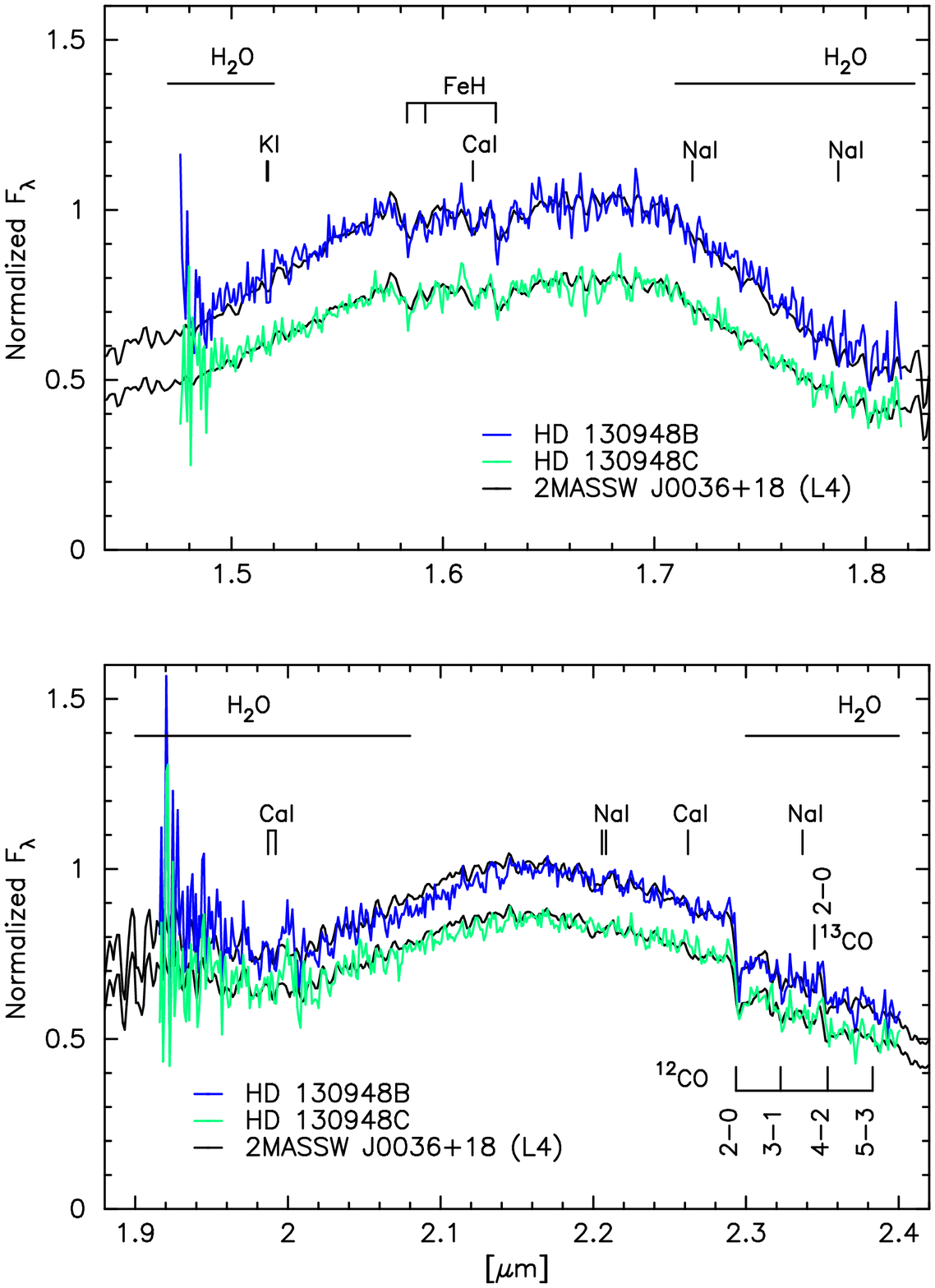}
\caption{}
\end{figure}


\figurenum{3}
\begin{figure}
\plotone{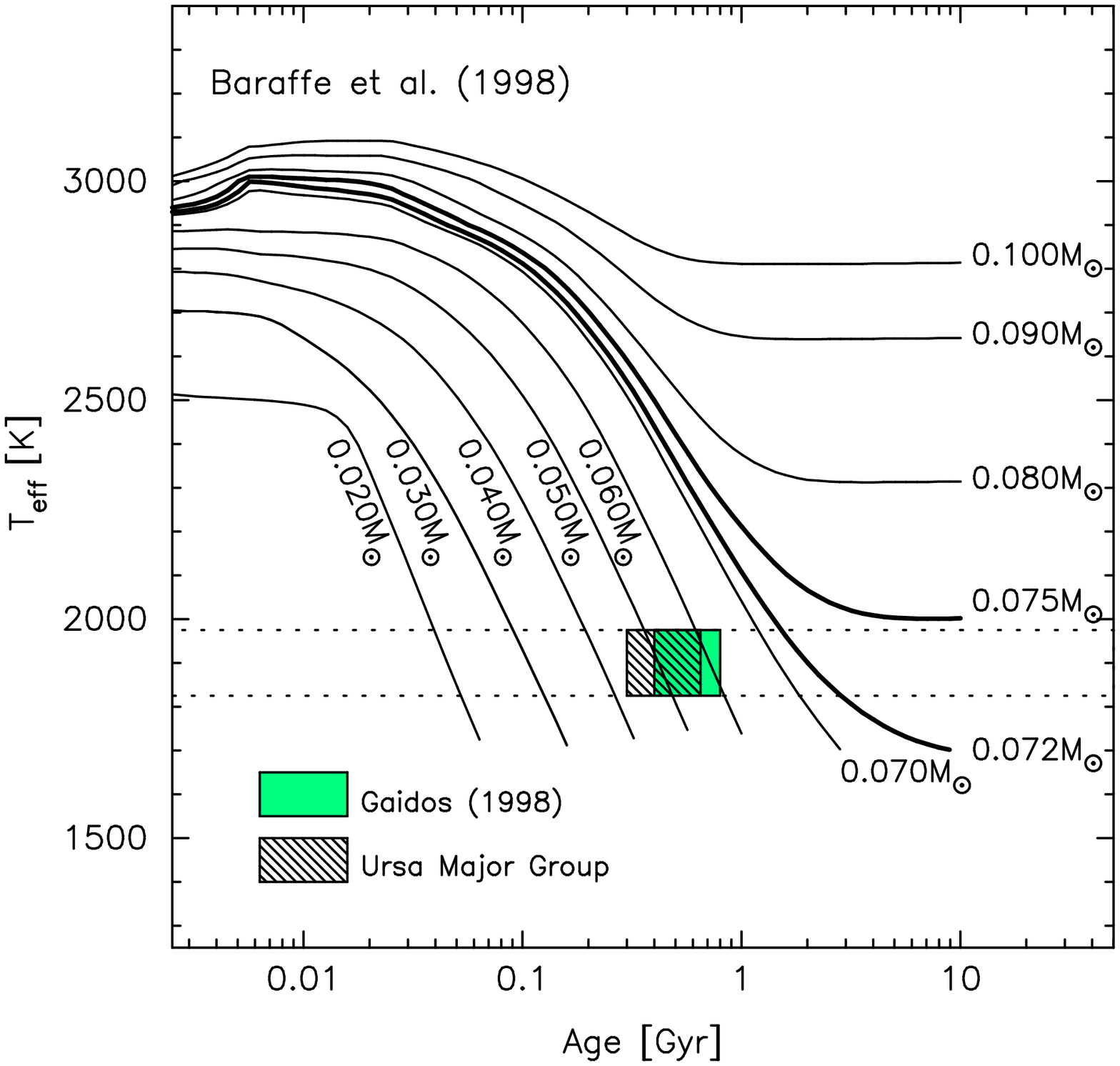}
\caption{}
\end{figure}

\end{document}